\documentclass[aps,pra,twocolumn,superscriptaddress,10pt,longbibliography]{revtex4-2}

\usepackage{color}
\usepackage{amsmath}
\usepackage{amssymb}
\usepackage{bm}
\usepackage{dsfont}
\usepackage[normalem]{ulem}
\usepackage{float}

\usepackage[T1]{fontenc}

\usepackage{circledsteps}
\usepackage{hyperref}
\usepackage[capitalize]{cleveref}
\crefrangeformat{figure}{Figs. #3#1#4--#5#2#6}
\crefrangeformat{equation}{Eqs. #3(#1)#4--#5(#2)#6}
\crefformat{section}{Sec. #2#1#3}

\usepackage{graphicx}
\graphicspath{{figs/}}

\newcommand{\pdagger}{{\vphantom{\dagger}}}
\newcommand{\pprime}{{\vphantom{\prime}}}

\newcommand{\beq}{\begin{eqnarray}}
\newcommand{\eeq}{\end{eqnarray}} 

\newcommand{\hide}[1]{}

\hypersetup{
  colorlinks=true,
  allcolors = blue,
  pdftitle={}
}

\newcommand{\figref}[2]{\namecref{#1}~\hyperref[#1]{\ref*{#1}(#2)}}

\begin{document}
	
\title{Expansion dynamics of strongly correlated lattice bosons: a selfconsistent density-matrix approach}

 	\author{Julian Schwingel}
	
	\email{s6jlschw@uni-bonn.de}

        \affiliation{Physikalisches Institut, Rheinische Friedrich-Wilhelms-Universit\"at Bonn, Nussallee 12, 53115, Bonn, Germany}

  	\author{Michael Turaev}
	
	\email{mturaev@uni-bonn.de}

	\affiliation{Physikalisches Institut, Rheinische Friedrich-Wilhelms-Universit\"at Bonn, Nussallee 12, 53115, Bonn, Germany}
    
	\author{Johann Kroha}
	
	\email{jkroha@uni-bonn.de}

        \affiliation{Physikalisches Institut, Rheinische Friedrich-Wilhelms-Universit\"at Bonn, Nussallee 12, 53115, Bonn, Germany}

        \affiliation{\hbox{School of Physics and Astronomy, University of St. Andrews, North Haugh, St. Andrews, KY16 9SS, United Kingdom}}

  	\author{Sayak Ray}
	
	\email{sayak@uni-bonn.de}

	\affiliation{Physikalisches Institut, Rheinische Friedrich-Wilhelms-Universit\"at Bonn, Nussallee 12, 53115, Bonn, Germany}

        \begin{abstract}
            We study the spatio-temporal dynamics of interacting bosons on a two-dimensional Hubbard lattice in the strongly interacting regime, taking into account the dynamics of condensate amplitude as well as the direct transport of non-condensed fluctuations. To that end we develop a selfconsistent density-matrix approach which goes beyond the standard Gutzwiller mean-field theory. Starting from the Liouville-von-Neumann equation we derive a quantum master equation for the time evolution of the system's local density matrix at each lattice site, with a dynamical bath that represents the rest of the system. We apply this method to  the expansion dynamics of an initially prepared cloud of interacting bosons in an optical lattice. We observe a ballistic expansion of the condensate, as expected, followed by slow, diffusive transport of the normal bosons. We discuss, in particular, the robustness of the Mott insulator phase as well as its melting due to incoherent transport. The method should be applicable to various models of lattice bosons in the strongly correlated regime.
        \end{abstract}
	
	
	\maketitle

\section{Introduction}

The experimental realization of the superfluid-to-Mott insulator (SF-MI) transition in the Bose-Hubbard model (BHM) opened up new avenues to study non-equilibrium dynamics of strongly interacting Bose gases \cite{Greiner02}. Since then, a variety of correlated models have been realized in the cold atom platform \cite{Bloch_RMP}. Moreover, the invention of quantum-gas microscopes has enabled the investigation of the spatio-temporal evolution of lattice bosons and their correlation properties in different dimensions as well as for different interaction strengths \cite{Greiner_2009, Greiner_2010}. The experimental observation of bimodal expansion of interacting Bose gases with ballistic and diffusive components \cite{Ronzheimer_2013, Kuhr_2012,Ronzheimer_2013} calls for a detailed understanding of these phenomena, involving the spatio-temporal interplay between the phase-coherent condensate dynamics and the incoherent excitations, especially in the strongly correlated regime. 

The quench dynamics of Bose-Hubbard systems in one dimension (1D) are well described by, for example, the time-dependent density matrix renormalization group (t-DMRG) and its recent variant, the matrix product states approach \cite{White_2004, Kollath_2007, Schollwoeck_2011}. Time-dependent exact diagonalization (ED) \cite{Kollath_2007} and variational quantum Monte Carlo (t-VMC) \cite{Gartner_2022} techniques are limited by short evolution times and finite system size and, thus, the difficulty to describe spontaneous symmetry breaking. However, due to the absence of long-range order in 1D a true condensation cannot occur \cite{Wagner_1966}. While true symmetry-breaking with the formation of a condensate can occur in two and higher dimensional systems, describing the spatio-temporal dynamics of those systems has remained a challenge. Selfconsistent, perturbative quantum field-theoretical methods were developed for the homogeneous temporal dynamics \cite{Rey_2004} and to capture the coupled dynamics of condensate and non-condensed fluctuations in the weakly interacting regime \cite{Trujillo-Martinez_2009,Lappe_2018,Trujillo-Martinez_2021}, which does not capture the Mott phase.  
The strongly interacting regime, on the other hand, is commonly described by time-dependent Gutzwiller methods based on single-site \cite{Hofstetter_07, Rigol_2011, Rigol17} or cluster mean-field theory \cite{Fazio_2016, Ray24}. This type of mean-field theory, however, neglects quantum hopping to neighboring sites and, thus, cannot describe temporal expansion initiating from a Mott localized state. In bosonic dynamical mean-field theory (B-DMFT), such hopping processes are included at second order \cite{Vollhardt_2008}. It has been applied to describe the non-equilibrium dynamics of Bose-Hubbard systems \cite{Eckstein15}, although only for spatially homogeneous conditions imposed by the DMFT assumption. 

In this work, we develop a method that is capable of describing the spatio-temporal dynamics of condensed as well as non-condensed bosons in the case of strong Hubbard repulsion and, ideally, near the Mott-superfluid transition. We derive the equations of motion for the density matrix at a local site from the corresponding von Neumann equation, with second-order hopping processes included in the fluctuation correlations with the neighboring sites. Approximations leading to a time-local form of these correlators make it possible to identify these equations as time-dependent quantum master equations for the local density matrix in space. This approach allows us to study the expansion dynamics of interacting bosons in an optical lattice described by the BHM. Starting from a realistic initial spatial distribution of atoms with a Mott insulator at its core, surrounded by a superfluid ring, we observe the spatio-temporal dynamics to display a clear separation between a ballistically expanding condensate and a slow, diffusively expanding Mott insulator. These results are consistent with the experiment in Ref.~\cite{Ronzheimer_2013}, which cannot be reproduced by the standard Gutzwiller mean-field methods \cite{Hofstetter_07, Rigol_2011}. The ballistic expansion velocity meets the Lieb-Robinson bound on information propagation \cite{Lieb_1972}, while the diffusive dynamics are corroborated by a slow expansion of the non-condensed atom cloud $\propto \sqrt{t}$. Furthermore, we compute the von Neumann number entropy at local sites, which supports the expansion behavior and is relevant for the experiments with quantum gas microscopes \cite{Greiner_2019}.

The paper is organized as follows. In \cref{sec:method} we develop the method and derive the explicit form of the master equation for the density matrix of the BHM in \cref{sec:BHM,sec:QME}. The different expansion dynamics of condensate and non-condensed cloud are demonstrated in \cref{sec:Expansion-dynamics}, starting with the ground state of a lattice Bose gas in a harmonic trap, consisting of a Mott localized core surrounded by a condensed ring, as a realistic initial state of the expansion. Subsequently, we analyze in \cref{sec:Mott-melting} the diffusive expansion of an initially purely Mott localized state and in \cref{sec:SF-expansion} the ballistic expansion of an initially completely condensed gas. We conclude in \cref{sec:conclusion} with a brief summary of the merits and limitations, and possible future applications of the present method and of the specific results for the two-dimensional Bose-Hubbard model.

\section{Model and method}
\label{sec:method}

\subsection{Density matrix approach to system dynamics coupled to a bath}

The B-DMFT is not directly applicable to the expansion scenario, as the expanding cloud does not obey spatial translation symmetry. We do, however, adopt the DMFT concept of isolating a single site (the impurity) and embedding it in an effective bath that represents the remaining sites. The temporal dynamics of the isolated site will be treated accurately, assuming that the condensate and non-condensed densities of the surrounding sites are known. In the case of spatio-temporal expansion, however, these quantities will evolve in time and will have to be determined selfconsistently by treating each site as the dynamical impurity in the above sense.  
In this spirit, we first set up our method for a generic class of models with system-bath interaction, which can be described by the following Hamiltonian:
\begin{equation}
    \hat{H} = \hat{H}_{\rm S} + \hat{H}_{\rm B} + \hat{H}_{\rm SB} \, ,
    \label{H-imp}
\end{equation}
where $\hat{H}_{\rm S,\, B}$ describes the system and the bath, respectively, and $\hat{H}_{\rm SB}$ the hybridization, which consists of the terms that act on both subsystems. The corresponding Liouville–von Neumann equation describing the dynamics of the system's density matrix is given by
\begin{equation}
    \partial_t \hat{\rho}(t) = -i \left[\hat{H}, \hat{\rho}(t)\right] .
\end{equation}
Note that here and throughout the paper, we have set $\hbar=1$. A Gutzwiller-like mean-field approximation would allow for inter-site hopping only if there is a condensate amplitude on the neighboring sites and would, therefore, never delocalize an initial Mott state. To allow for transport of non-condensed particles by the hybridization Hamiltonian, we now switch to the interaction picture where the time evolution of an operator $\hat{O}$ from time $t_0$ to $t$ is governed by $\hat{O}^I(t) = \hat{U}^\dagger(t,t_0) \hat{O}(t_0) \hat{U}(t,t_0)$ where $\hat{U}(t,t_0)=\hat{U}_{\rm B}(t,t_0) \hat{U}_{\rm S}(t,t_0)$, with the unitary time-evolution operators in the system, $\hat{U}_{\rm S}(t,t_0)$, and in the bath, $\hat{U}_{\rm B}(t,t_0)$, defined as
\begin{equation}
    \hat{U}_{\rm S,\,B}(t,t_0) = \hat{\mathcal{T}} e^{-i \int_{t_0}^t \mathrm{d}t' \hat{H}_{\rm S,\,B}(t')} \, ,
    \label{eq:time_evolution}
\end{equation}
and the time-ordering operator $\hat{\mathcal{T}}$. Note that $\hat{H}_{\rm S}$ and $\hat{H}_{\rm B}$ of the isolated system and bath are \textit{a priori} not time dependent, but will acquire an effective time dependence due to the dynamical partitioning into condensate and non-condensate particle densities, see below, which leads to the time-ordered integral in \cref{eq:time_evolution}. The von Neumann equation for the density matrix $\hat{\rho}^I(t)$ in the interaction picture is given by
\begin{equation}
\label{eq:vonNeumannInt}
    \partial_t \hat{\rho}^I(t) = -i [\hat{H}_{\rm SB}^I(t), \hat{\rho}^I(t)]\, ,
\end{equation}
which reads in integral form
\begin{equation}
    \hat{\rho}^I(t) = \hat{\rho}(t_0) - i \int_{t_0}^t dt' \, [\hat{H}_{\rm SB}^I(t'), \hat{\rho}^I(t')]\, .
    \label{eq:rhot-int}
\end{equation}
By plugging \cref{eq:rhot-int} into \cref{eq:vonNeumannInt}, followed by tracing over the bath, we obtain the master equation for the system density-matrix to second order in the system-bath coupling, 
\begin{equation}
    \partial_t \hat{\rho}^I_{\rm S}(t) = - \int_{t_0}^t \mathrm{d}t' ~ {\rm Tr_B}\left[\hat{H}_{\rm SB}^I(t),\left[\hat{H}_{\rm SB}^I(t'),\hat{\rho}^I(t')\right]\right], 
    \label{eq:EOMrho0}
\end{equation}
where it is assumed that the system-bath coupling vanishes for $t\leq t_0$, so that ${\rm Tr_B}\left[\hat{H}_{\rm SB}^I(t),\hat{\rho}^I(t_0)\right]=0$ \cite{breuer-book}.
Finally, by performing an inverse transformation of \cref{eq:EOMrho0} from the interaction picture, the time-evolved system density matrix $\hat{\rho}_{\rm S}(t)$ in the Schr\"odinger picture is obtained, and the dynamics of the observables in the system can be studied.

Starting from \cref{eq:EOMrho0}, we will now apply this density matrix approach to study the spatio-temporal dynamics of the BHM as a generic example of interacting lattice-boson models. 

\subsection{Application to the Bose-Hubbard model}
\label{sec:BHM}

The BHM is described by the Hamiltonian \cite{Fisher_1989, Jaksch_1998},
\begin{align}
\hat{\mathcal{H}}_{\rm BHM} &= -J\sum_{\langle {\bf r},{\bf r^{\prime}}\rangle} \hat{b}_{\bf r^\pprime}^{\dagger}\hat{b}_{\bf r^{\prime}}^\pdagger + \sum_{\bf r} \hat{H}_{\bf r}^{\rm loc} \nonumber \\ 
{\rm with} ~ ~ \hat{H}_{\bf r}^{\rm loc} &= \left[\frac{U}{2} \hat{n}_{\bf r}(\hat{n}_{\bf r}-1) + V_{\bf r} \hat{n}_{\bf r} \right],
\label{BHM}
\end{align}
where $\hat{b}_{\bf r}^\pdagger$, $\hat{b}_{\bf r}^{\dagger}$ and $\hat{n}_{\bf r}$ are the bosonic annihilation, creation, and number operators, respectively, at site ${\bf r} = (i_x,i_y)$ of a square lattice of size $L \times L$. $J$ is the nearest-neighbor hopping amplitude, $U$ the on-site interaction strength, and $V_{\bf r}$ an external potential, which can arise due to a magneto-optical trap or disorder, for example.

\begin{figure}[t]
    \includegraphics[width=1.0 \columnwidth]{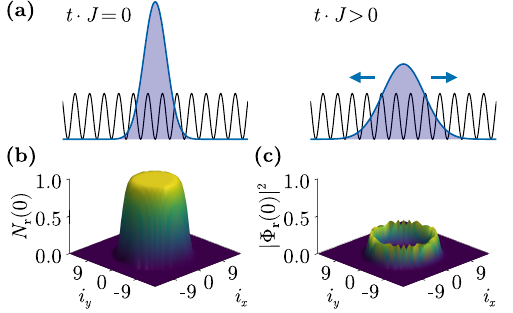}
    \caption{{\it Schematic of the setup.} (a) Sketch of the expansion of an interacting, bosonic cloud in an optical lattice. (b) and (c) show the zoomed-in spatial profiles of the particle density $N_{\bf r}(0)$ and of the superfluid density $|\Phi_{\bf r}(0)|^2$, respectively, for an initial Mott insulating core surrounded by a superfluid ring. The initial profile is computed using the Gutzwiller mean-field theory for $U/J=50$, $V_0/J=0.2$ and chemical potential $\mu/J=15$, see the text.}
    \label{fig:schematics}
\end{figure}

The Hamiltonian, \cref{BHM}, can be partitioned analogous to \cref{H-imp} as $\hat{\mathcal{H}}_{\rm BHM}=\hat{H}_{\rm S}+\hat{H}_{\rm B}+\hat{H}_{\rm SB}$, with
\begin{align}
\begin{split}
    \hat{H}_{\rm S} = \hat{H}_{\bf r}^{\rm loc} , \quad
    \hat{H}_{\rm B} &= -J\sum_{\langle {\bf r'},{\bf r''}\rangle \in ({\bf r})} \hat{b}_{\bf r'}^{\dagger}\hat{b}_{\bf r''}^\pdagger + \sum_{{\bf r'}\neq {\bf r}} \hat{H}_{\bf r'}^{\rm loc} ,
    \\
    \hat{H}_{\rm SB} &= -J \sum_{{\bf r'} \in {\bf R_r}} \hat{b}_{\bf r'}^\dagger \hat{b}_{\bf r^\pprime}^\pdagger + {\rm H.c.} \, , \label{eq:cavity-H}
\end{split}
\end{align}
where an arbitrarily chosen site ${\bf r}$, which is fixed for the moment, represents the system and is coupled to the other lattice sites serving as a bath. Here, $({\bf r})$ denotes the set of all lattice sites except ${\bf r}$, and ${\bf R_r}$ the set of nearest neighbor sites of ${\bf r}$ \cite{Vollhardt_2008}.
By defining the condensate amplitude at site ${\bf r}$ as $\Phi_{\bf r}(t) = \langle \hat{b}_{\bf r} \rangle$ and the fluctuations as $\delta\hat{b}_{\bf r}(t) = \hat{b}_{\bf r} - \Phi_{\bf r}(t)$, the hopping term can be decomposed into mean-field (MF) parts in the system (S) and on the neighboring sites (B) and a fluctuation part (fl) as, $\hat{H}_{\rm SB} = \hat{H}_{\rm S}^{\rm MF}(t) + \hat{H}_{\rm B}^{\rm MF}(t) + \hat{H}_{\rm SB}^{\rm fl}(t)$, with
\begin{align}
    \hat{H}_{\rm S}^{\rm MF}(t) &= -J \Phi_{\rm eff,{\bf r}}^{*}(t) \hat{b}_{\bf r}^\pdagger + {\rm H.c.} \, , \nonumber\\ 
    \hat{H}_{\rm B}^{\rm MF}(t) &= - J \sum_{{\bf r'} \in {\bf R_r}} \Phi_{\bf r^\pprime}^\pdagger(t) \hat{b}_{\bf r'}^{\dagger} + {\rm H.c.} \label{eq:H-hybrid} \, , \\ 
    \hat{H}_{\rm SB}^{\rm fl}(t) &= -J \sum_{{\bf r'} \in {\bf R_r}} \delta\hat{b}_{\bf r'}^\dagger(t) \delta\hat{b}_{\bf r^\pprime}^\pdagger(t) + {\rm H.c.} \, , \nonumber
\end{align}
where $\Phi_{\rm eff,\, {\bf r}}(t)=\sum_{\substack{{\bf r'} \in {\bf R_r}}} \Phi_{\bf r'}(t)$ is the sum of the condensate amplitudes on the neighboring sites. Since $\hat{H}_{\rm S,\,B}^{\rm MF}(t)$ are local operators in the system or the bath, respectively, we can absorb these mean-field contributions in an effective system or bath Hamiltonian, respectively, and write the time-evolution operators of the uncoupled system and bath as (cf. Eq.~\eqref{eq:time_evolution})
\begin{equation}
    \hat{U}_{\rm S,\,B}(t,t_0) = \hat{\mathcal{T}} e^{-i \int_{t_0}^t \mathrm{d}t' \left(\hat{H}_{\rm S,\,B} + \hat{H}_{\rm S,\,B}^{\rm MF}(t')\right)}
    \label{eq:Unitary-interaction}
\end{equation}
Note that $\hat{H}_{\rm S} + \hat{H}_{\rm S}^{\rm MF}(t)$ corresponds to the Gutzwiller mean-field Hamiltonian \cite{Rigol_2011}, such that its dynamics are contained in the operator $\hat{U}_{\rm S}(t,t_0)$. Following \cref{eq:EOMrho0}, the density matrix $\hat{\rho}_{\bf r}^I(t)$ of a local site ${\bf r}$ is time evolved with respect to $\hat{H}_{\rm SB}^{{\rm fl},\,I}(t)$ in the interaction picture, with its equation of motion given by
\begin{equation}
    \partial_t \hat{\rho}^I_{\bf r}(t) = - \int_{t_0}^t \mathrm{d}t' ~ \mathrm{Tr}_{({\bf r})}\left[\hat{H}_{\rm SB}^{{\rm fl},\,I}(t),\left[\hat{H}_{\rm SB}^{{\rm fl},\,I}(t'),\hat{\rho}^I(t')\right]\right].
    \label{eq:EOMrho0-BHM}
\end{equation}
We note in passing that the above partitioning is analogous to the cavity construction of DMFT, where one derives an effective action for the local (impurity) site after tracing over the bath sites, with the assumption that the self-energy is local \cite{Vollhardt_2008}. 
In a similar spirit, spatial quantum correlations (beyond mean field) are neglected in our method, resulting in the factorization of the total density matrix as, $\hat{\rho}(t)=\prod_{\bf r} \hat{\rho}_{\bf r}(t)$. Dropping the fluctuation term $\hat{H}_{\rm SB}^{\rm fl}(t)$ in \cref{eq:H-hybrid} would, thus, be equivalent to the single-site Gutzwiller mean-field theory \cite{Hofstetter_07, Rigol_2011}. 

\subsection{Derivation of the quantum master equation}
\label{sec:QME}

Following \cref{eq:EOMrho0}, we now proceed to derive the time-evolution equation for the density matrix $\hat{\rho}_{\bf r}(t)$ at site ${\bf r}$ of the BHM.  
For convenience, we represent the field operators, $\hat{b}_{\bf r}^\pdagger$, $\hat{b}_{\bf r}^{\dagger}$, and the condensate amplitudes, $\Phi_{\bf r}^{\vphantom{*}}$, $\Phi_{\bf r}^*$, in Nambu space as 
\begin{equation}
    \boldsymbol{b}_{\bf r} = 
    \begin{pmatrix}
        \hat{b}_{\bf r}^\pdagger \\
        \hat{b}_{\bf r}^\dagger
    \end{pmatrix},  \quad \boldsymbol{\varphi}_{\bf r} =     \begin{pmatrix}
        \Phi_{\bf r}^{\vphantom{*}} \\
        \Phi_{\bf r}^*
    \end{pmatrix},
\end{equation}
respectively. 
The master equation for the density matrix (cf. \cref{eq:EOMrho0-BHM}) can then be written as
\begin{align}
    \partial_t \hat{\rho}^I_{\bf r}(t) &= -J^2 \int_{t_0}^t \mathrm{d}\bar{t} \sum_{{\bf r'},{\bf r''} \in {\bf R_r}} \sum_{\alpha,\, \beta} \left( \left[b^{\dagger \, I}_{{\bf r^\pprime},\,\alpha}(t), b^{I\pdagger}_{{\bf r},\, \beta}(\bar{t}) \hat{\rho}^I_{\bf r}(\bar{t})\right] \right. \nonumber \\
    &- \left. \varphi_{{\bf r^\pprime},\,\beta}(\bar{t}) \left[b^{\dagger \, I}_{{\bf r^\pprime},\, \alpha}(t), \hat{\rho}^I_{\bf r}(\bar{t})\right] \right) G^{(c) \alpha,\,\beta}_{{\bf r',r''}}(t,\bar{t}) + {\rm H.c.} \, ,
    \label{eq:EOM-rho0-BHM_1}
\end{align}
where $\alpha$ and $\beta$ denote the Nambu indices, and $G^{(c) \alpha,\,\beta}_{{\bf r',r''}}(t,\bar{t})$ are the components of the connected part of the two-time correlator $\boldsymbol{G}_{{\bf r'},\,{\bf r''}}^{(c)}(t,\bar{t})$ defined as
\begin{equation}
   G_{{\bf r'},\,{\bf r''}}^{(c) \alpha,\, \beta}(t,\bar{t}) = \mathrm{Tr}_{({\bf r})}\left\{ {\delta}b^{I\pdagger}_{{\bf r'},\, \alpha}(t) \delta b^{\dagger\, I}_{{\bf r''},\, \beta}(\bar{t}) \hat{\rho}_{({\bf r})}^{I}(\bar{t}) \right\} \, . 
   \label{eq:2-time-corr}
\end{equation}
We note that solving an integro-differential equation like \cref{eq:EOM-rho0-BHM_1} is challenging. To perform the retarded time integral, an exponential form of the bath correlation function has previously been assumed, see, for instance, Ref.~\cite{Strunz_2017}. 
In a similar spirit, we re-write \cref{eq:EOM-rho0-BHM_1} in Wigner relative and center-of-motion (CoM) time coordinates, $\tau=t-\bar{t}$ and $T=(t+\bar{t})/2$, respectively. For the present case (short-time correlations due to strong onsite repulsion $U$), we assume that in Wigner coordinates the two-time correlator is a sharply peaked function at time difference $\tau=0$ with an exponential decay in $\tau$, 
\begin{equation}
\boldsymbol{G}_{{\bf r'},\,{\bf r''}}^{(c)}(t,\bar{t}) \rightarrow \boldsymbol{G}_{{\bf r'},\,{\bf r''}}^{(c)}(T,\tau) = \boldsymbol{G}_{{\bf r'},\,{\bf r''}}^{(c)}(T,\tau=0)\,\mathrm{e}^{-\tau/\tau_{\rm corr}} ,
\label{eq:EqualTimeApprox}
\end{equation}
with the correlation time $\tau_{\rm corr}\simeq 1/U$, while the rest of the integrand in \cref{eq:EOM-rho0-BHM_1} varies slowly as function of $\tau$, on a scale much longer than $\tau_{\rm corr}$. It can, thus, be pulled out of the $\tau$-integral and evaluated at equal times, $\tau=0$, so that the remaining part of the $\tau$-integral can be performed.  
Furthermore, since the density matrix factorizes in space, we obtain $\boldsymbol{G}_{{\bf r'},\,{\bf r''}}^{(c)}(T,\tau=0):=\boldsymbol{G}_{{\bf r'}}^{(c)}(t)\delta_{{\bf r'},{\bf r''}}$. Transforming back to the Schr\"odinger picture re-introduces the on-site, coherent dynamics of the system by means of a von Neumann commutator (first term in \cref{eq:EOMrho0Markovian}), and we arrive at the master equation for $\hat{\rho}_{\bf r}(t)$ at the CoM time $t = T$ and $\tau=0$,
\begin{equation}
    \partial_t \hat{\rho}_{\bf r}(t) = 
    - i [\hat{H}_{\rm S} + \hat{\widetilde{H}}\vphantom{\hat{H}}_{\rm S}^{\rm MF}(t), \hat{\rho}_{\bf r}(t)] + J^2 \tau_{\rm corr} \mathcal{L}[\boldsymbol{b}_{\bf r}^\pdagger] \, . 
    \label{eq:EOMrho0Markovian}
\end{equation}
The last term in this equation reads, 
\begin{align}
    \mathcal{L}[\boldsymbol{b}_{\bf r}^\pdagger]
    &= \sum_{{\bf r'} \in {\bf R_r}} \
    \sum_{\alpha,\beta} {G}^{(c)\alpha,\beta}_{\bf r'}(t) \left[ 2 {b}_{{\bf r}, \beta}^{\vphantom{*}} \hat{\rho}_{\bf r}(t) {b}^{\dagger}_{{\bf r},\alpha} \right. \nonumber \\
    &- \left. {b}^{\dagger}_{{\bf r}, \alpha} {b}_{{\bf r}, \beta}^{\vphantom{*}} \hat{\rho}_{\bf r}(t) - \hat{\rho}_{\bf r}(t) {b}^{\dagger}_{{\bf r}, \alpha} {b}_{{\bf r}, \beta}^{\vphantom{*}} \right]\, , 
    \label{eq:D-fluctuation}
\end{align}
with the local, equal-time correlator,
\begin{align}
    {G}_{\bf r'}^{(c) \alpha, \beta}(t) =& \mathrm{Tr}\left\{ {\delta}{b}_{{\bf r'}, \alpha}^\pdagger(t) \delta {b}^{\dagger}_{{\bf r'}, \beta}(t) \hat{\rho}_{{\bf r}}(t) \right\} \label{eq:Eq-time-corr} \\
    =& \mathrm{Tr} \left\{ {b}_{{\bf r'}, \alpha}^\pdagger {b}^{\dagger}_{{\bf r'}, \beta} \hat{\rho}_{{\bf r'}}(t) \right\} - {\varphi}_{{\bf r'}, \alpha}^{\vphantom{*}}(t) {\varphi}^*_{{\bf r'}, \beta}(t) \nonumber \, ,
\end{align}
where the trace runs over the local Fock space at the nearest-neighbor site ${\bf r'}$.
The local system Hamiltonian which is renormalized by the Lamb-shift term (first term on the right-hand side of \cref{eq:EOMrho0Markovian}) is given by,
\begin{equation}
    \hat{\widetilde{H}}\vphantom{\hat{H}}_S^{\rm MF}(t) = \hat{H}_S^{\rm MF}(t) + i J^2 \tau_{\rm corr} \sum_{{\bf r'} \in {\bf R_r}} \left(\Phi_{\bf r'}^\pdagger(t) \hat{b}_{\bf r}^\dagger - \Phi_{\bf r'}^{*\pdagger}(t) \hat{b}_{\bf r}^\pdagger\right).
    \label{eq:Lamb-Shift}
\end{equation}
The detailed derivation to arrive at this form of the Master equation is discussed in \cref{deriv-QME}.
Once the density matrix $\hat{\rho}_{\rm r}(t)$ is obtained, one can compute any local, single-time expectation value as, 
\begin{align}
\langle (\dots) \rangle = {\rm Tr}\left[ \hat{\rho}_{\bf r}(t) (\dots)\right]\,, 
\label{eq:exp_values}
\end{align}
for instance, the local particle number, $N_{\bf r}(t):=\langle \hat{n}_{\bf r}\rangle$, the condensate amplitude, $\Phi_{\bf r}(t):=\langle \hat{b}_{\bf r}\rangle$, the non-condensed number, $N_{\bf r}^{\rm fl}(t)=N_{\bf r}(t)-|\Phi_{\bf r}(t)|^2$, and, in particular, the spatially and temporally local propagator ${\bf G}_{\bf r'}^{(c)}(t)$ appearing on the right-hand side of \cref{eq:EOMrho0Markovian}. 
\begin{figure}[t]
    \includegraphics[width=\linewidth]{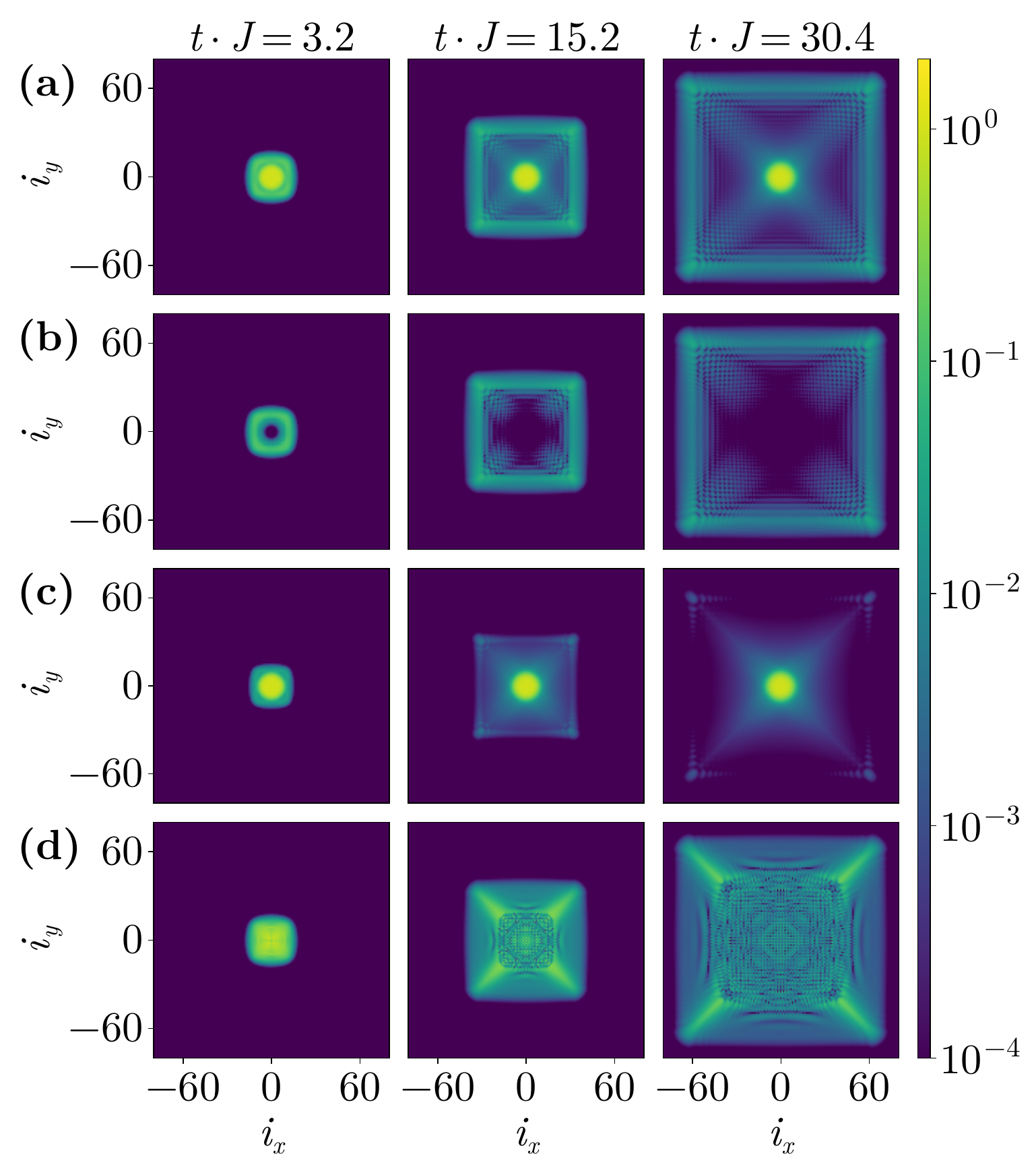}
    \caption{{\it Density-matrix vs. Gutzwiller mean-field theory.} Spatial profiles of (a) particle-number density $N_{\bf r}(t)$, (b) condensate density $|\Phi_{\bf r}(t)|^2$ and (c) non-condensed density $N_{\bf r}^{\rm fl}(t)$ at different times (see at the top) are shown as a color scale in a square lattice using the density matrix approach for $U/J=50$. The particle-number densities $N_{\bf r}(t)$ obtained using the Gutzwiller mean-field method are shown in (d) for comparison.}
    \label{fig:GW-DM-Profiles}
\end{figure}
Thus, \crefrange{eq:EOMrho0Markovian}{eq:exp_values} constitute a closed set of coupled, first-order-in-time differential equations for the time evolution of $\hat{\rho}_{\bf r}(t)$ on each lattice site. 
The numerical time evolution is performed using an adaptive Runge-Kutta time-stepping algorithm, where at each instant of time $t$ the right-hand side of \cref{eq:EOMrho0Markovian} is computed using the values of $\hat{\rho}_{\bf r}(t-\delta t)$, $\Phi_{\bf r}(t-\delta t)$ and ${\bf G}_{\bf r}^{(c)}(t-\delta t)$ known from the previous time steps. 
All the operators are written in the local Fock state basis $|n\rangle$ ($n=0,1,2,\cdots$) where the local density matrix $\hat{\rho}_{\bf r}(t)$ reads,
\begin{equation}
    \hat{\rho}_{\bf r}(t) = \sum_{n,m =0}^{\infty} \rho_{\bf r}^{n,m}(t) |n\rangle \langle m|\, .
\end{equation}
Here, $\rho_{\bf r}^{n,n}(t)=\langle n | \hat{\rho}_{\bf r}(t) | n \rangle$ is the time-dependent probability for the occupation number $n$ on site ${\bf r}$. We note that our master equation approach is trace preserving, that is, the total Fock state probability in $\hat{\rho}_{\bf r}$ is conserved, $\sum_n \rho_{\bf r}^{n,n}(t)=1$, see \cref{app:B}, and yields that the time-evolved density matrix $\hat{\rho}_{\bf r}(t)$ is always positive definite. Moreover, the total particle number $N_{\rm tot}=\sum_{\bf r}N_{\bf r}(t)$ is conserved, and is set by the initial density distributions.
For the numerical calculations, we truncate the local Fock space at $n_{\rm max}$ such that the probability corresponding to the highest occupation state $\rho_{\bf r}^{n_{\rm max},n_{\rm max}}(t)$ remains $\lesssim 10^{-5}$ which is achieved typically for $n_{\rm max}\gtrsim 15$. The numerical time evolutions are performed on a homogeneous ($V_{\bf r}=0$), $L\times L$ square lattice with $L=161$ and for correlation time $\tau_{\rm corr}\cdot U=1$.

In comparison to a space and time dependent B-DMFT, the density-matrix approach provides a numerically inexpensive method to calculate the time dependence of occupation numbers, amplitudes and any higher order correlators local in time and space, which are the most important quantities for spatio-temporal evolution in typical cold-atom experiments. While nonlocal-in-time quantities have been computed using nonequilibrium B-DMFT for spatially homogeneous cases \cite{Eckstein15}, our method can reach orders of magnitude larger timescales, which is important for inhomogeneous expansion dynamics.

\begin{figure}[t]
    \includegraphics[width=\linewidth]{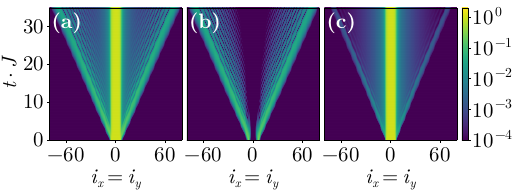}
    \caption{{\it Spatio-temporal evolution of atomic densities.} The spatio-temporal dynamics of (a) the particle-number density $N_{\bf r}(t)$, (b) condensate density $|\Phi_{\bf r}(t)|^2$ and (c) non-condensed density $N_{\bf r}^{\rm fl}(t)$ corresponding to \figref{fig:GW-DM-Profiles}{a--c}, respectively, are shown along the diagonal cut of the square lattice.}
    \label{fig:DM-Timeevol}
\end{figure}

\section{Expansion dynamics of bosons in Hubbard lattice}
\label{sec:Expansion-dynamics}

We now turn to the numerical evaluation of the dynamical equations for an expanding Bose gas in the BHM. Corresponding to the experiment Ref.~\cite{Ronzheimer_2013}, we prepare the system initially in the ground state of the BHM, \cref{BHM}, with an external harmonic trap potential $V_{\bf r}=V_0{\bf r}^2$, using the self-consistent Gutzwiller mean-field theory \cite{Rigol_2011}. We adjust the particle number in the trap center to 1 by means of a chemical potential such that the spatial distribution of atoms is a Mott insulating core with local occupation number 1 surrounded by a superfluid ring as shown in \cref{fig:schematics}. Since this is not an equilibrium state of the master equation (\ref{eq:EOMrho0Markovian}),  the initially prepared atomic cloud exhibits expansion dynamics in the homogeneous lattice ($V_{\bf r}=0$) shown by the spatial atom-density profiles $N_{\bf r}(t)$ at different time instants in \figref{fig:GW-DM-Profiles}{a}. The corresponding condensate and non-condensate density profiles are shown in \figref{fig:GW-DM-Profiles}{b--c}, and will be discussed in detail in the following sections. For comparison, we have also plotted in \figref{fig:GW-DM-Profiles}{d} the atom-density profiles $N_{\bf r}(t)=\langle \hat{n}_{\bf r}\rangle$ using the time-dependent Gutzwiller mean-field method as discussed previously in Ref.~\cite{Rigol_2011}. Within our density matrix approach, a square-shaped expansion of the atoms in the outer ring is observed, while the density profile of the core remains almost the same over the observed time frame, see \figref{fig:GW-DM-Profiles}{a}. The fast expansion of the outer wing, followed by a slow melting of the Mott insulating core, is consistent with the experiment \cite{Ronzheimer_2013}, which is not captured by the time-dependent Gutzwiller mean-field dynamics.

\subsection{Dynamics of condensate and non-condensed atom densities}
\label{sec:Expansion-cond-NC}

\begin{figure}[t]
    \includegraphics[width=\linewidth]{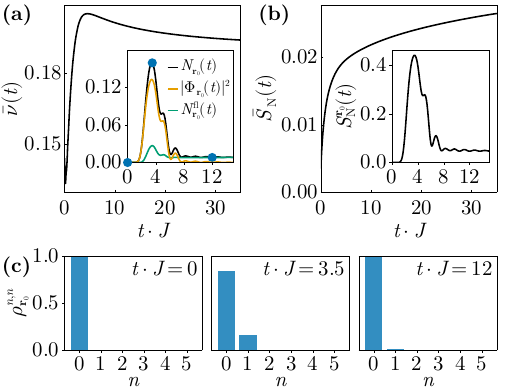}
    \caption{{\it Dynamics of condensate fraction and number entropy.} Time evolution of (a) the condensate fraction $\bar{\nu}(t)$ and (b) average number entropy $\bar{S}_{\rm N}(t)$ corresponding to \crefrange{fig:GW-DM-Profiles}{fig:DM-Timeevol} are shown. The insets show the representative dynamics of (a) the condensate density $|\Phi_{{\bf r}_0}(t)|^2$ and non-condensed density $N_{{\bf r}_0}^{\rm fl}(t)$, and (b) the number entropy $S_{\rm N}^{{\bf r}_0}(t)$ at ${\bf r}_0=(10,10)$. The corresponding Fock state distributions $\rho_{{\bf r}_0}^{n,n}(t)$ are shown in (c) at different time instants as given and marked by filled circles in (a) inset.}
    \label{fig:Evolution-local}
\end{figure}

The expansion dynamics are further investigated by plotting the spatial profiles of the condensate $|\Phi_{\bf r}(t)|^2$ and non-condensed densities $N_{\bf r}^{\rm fl}(t)$ in \figref{fig:GW-DM-Profiles}{b--c} and the corresponding spatio-temporal evolution in \cref{fig:DM-Timeevol}. The total particle number $N=\sum_{\bf r} N_{\bf r}(t)$ remains constant during time evolution, while the total condensate density exhibits an initial increase followed by a slow decrease during the expansion, see the time evolution of condensate fraction, $\bar{\nu}(t)=\sum_{\bf r} |\Phi_{\bf r}(t)|^2/N$, in \figref{fig:Evolution-local}{a}. Two distinctly different behaviors in the expansion of condensate and non-condensed particles are observed. The condensate atoms expand ballistically with velocity $2J$ along each side and $2\sqrt{2}J$ along the diagonals consistent with the Lieb-Robinson bound \cite{Lieb_1972}, see \cref{fig:DM-Timeevol}, leading to a square-shape condensate density profile as shown in \figref{fig:GW-DM-Profiles}{b}. Whereas the non-condensed atoms expand in two ways, see \figref{fig:GW-DM-Profiles}{c} and \cref{fig:DM-Timeevol}, as is explained below. 

Those non-condensed atoms generated from the condensate at local sites and, thus, dragged along with it, yield a spatio-temporal profile similar to that of the condensate atoms. An example of the dynamics of a local site ${\bf r}_0$ outside of the initial density profile is shown in the inset of \figref{fig:Evolution-local}{a}. Initially ($t\cdot J=0$) ${\bf r}_0$ is in a vacuum state, $\rho_{{\bf r}_0}^{n,m}(0)=\delta_{n,m}\delta_{n,0}$. As the fast, ballistically expanding condensate hits the site ($t\cdot J \approx 2$), the local condensate density and, consequently, the non-condensed atom density increases (cf. \figref{fig:Evolution-local}{a} inset). After the condensate leaves ($t\cdot J \approx 10$), the site is left with non-condensed atoms described by the diagonal, mixed density matrix with $\rho_{{\bf r}_0}^{n,m}(t)=\rho_{{\bf r}_0}^{n,n}(t)\delta_{n,m}$. The corresponding number state distributions $\rho_{{\bf r}_0}^{n,n}(t)$ are shown in \figref{fig:Evolution-local}{c}.
Surprisingly, the non-condensed atoms in the Mott insulating core, unlike in the Gutzwiller mean-field method \cite{Rigol_2011}, exhibit a robust structure and expand slowly for a long time due to their direct hopping in the lattice, see \cref{fig:DM-Timeevol} and \cref{sec:Mott-melting}. 

The dynamical behavior---a fast, ballistically expanding wing of superfluid condensate, and a slowly spreading Mott insulator core, which arises due to the interplay between the condensate and non-condensed particles in the strongly interacting regime---is the main result of this paper, and is consistent with the experimental observation \cite{Ronzheimer_2013}. The two qualitatively different expansion behaviors exhibited by the condensate and non-condensed atoms are further investigated in detail starting from specific initial distributions of atoms in sections \ref{sec:Mott-melting} and \ref{sec:SF-expansion}.

\subsection{Growth of von Neumann number entropy}
\label{sec:vN-number-entropy}

The expansion dynamics are corroborated by an increase in the average entropy of the system. In order to demonstrate this, we compute the site-averaged von Neumann number entropy $\bar{S}_{\rm N}(t)$ from the time-evolved density matrix $\hat{\rho}_{\bf r}(t)$ given by \cite{Greiner_2019} 
\begin{equation}
    \bar{S}_{\rm N}(t) = \frac{1}{L^2}\sum_{\bf r} S_{\rm N}^{\bf r}(t), ~~S_{\rm N}^{\bf r}(t) = -\sum_n \rho_{\bf r}^{n,n}(t) \ln \rho_{\bf r}^{n,n}(t),
\end{equation}
where $S_{\rm N}^{\bf r}(t)$ denotes the number entropy of a local site ${\bf r}$. The condensate fraction, averaged over the lattice sites, $\bar{\nu}(t)$ and the growth of $\bar{S}_{\rm N}(t)$ with time are shown in \figref{fig:Evolution-local}{a} and {\hyperref[fig:Evolution-local]{(b)}, respectively. At a site ${\bf r}_0$, the number entropy $S_{\rm N}^{{\bf r}_0}(t)$ behaves in the same way as the number densities (cf. \figref{fig:Evolution-local}{a,b}, insets), that is, the entropy grows from $0$ to a finite value as the cloud hits the site, followed by a decrease and then relaxation to a finite value in the non-condensed, low-density state as the cloud leaves the site (cf. \figref{fig:Evolution-local}{b} inset). We note that the number entropy is directly measurable and therefore particularly relevant for experiments using a quantum gas microscope \cite{Greiner_2019}. 
The von Neumann entropy of the local density matrix, $S_{\rm vN}^{\bf r}(t)=-{\rm Tr}[\hat{\rho}_{\bf r}(t)\ln\hat{\rho}_{\bf r}(t)]$, exhibits similar behavior, where $\rm Tr$ denotes the trace over the local Fock space.

\begin{figure}[t]
    \includegraphics[width=\linewidth]{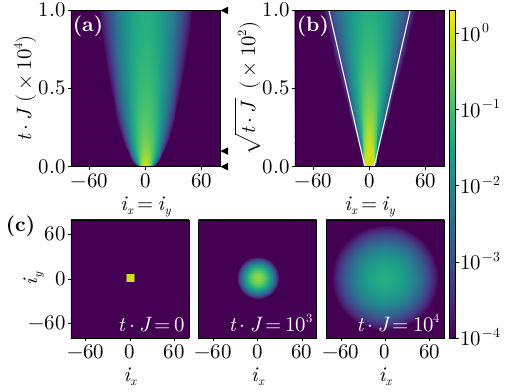}
    \caption{{\it Slow melting of Mott insulator.} The spatio-temporal dynamics of the particle-number density $N_{\bf r}(t)$ from an initially prepared Mott insulating core ($N_{\bf r}(0)=1$) of size $11\times 11$ are shown for $U/J=100$ along the diagonal cut of the square lattice in (a) and (b) with time plotted in linear and square-root scale, respectively. The straight, white lines in panel (b) are guides to the eye. The spatial density profiles are shown in panel (c) at different time instants as given and marked by arrowheads in (a).}
    \label{fig:DM-Timeevol-Mott}
\end{figure}

\section{Long-time behavior of the non-condensate dynamics}
\label{sec:Mott-melting}

\begin{figure}[t]
    \includegraphics[width=\linewidth]{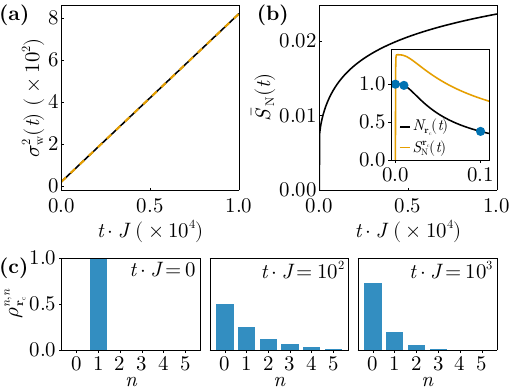}
    \caption{{\it Diffusion dynamics and number entropy.} (a) The width of the expanding cloud of non-condensed particles $\sigma_{\rm w}(t)$ shows diffusive growth. The dashed line is a linear fit to the $\sigma_{\rm w}^2(t)$ vs $t$ curve (solid line) for long times. (b) The average number entropy $\bar{S}_{n}(t)$ corresponding to \cref{fig:DM-Timeevol-Mott} is shown as function of time.  The inset shows the dynamics of particle-number density $N_{{\bf r}_{\rm c}}(t)$ and number entropy $S_{\rm N}^{{\bf r}_{\rm c}}(t)$ at the central site, ${\bf r}_{\rm c}=(0,0)$. The corresponding number-state distributions are shown in (c) at different times given and marked by filled circles in (b) inset .}
    \label{fig:S-and-width-Mott}
\end{figure}

To understand the long-time behavior of the slow dynamics exhibited by the non-condensed atoms, explicitly, due to their direct hopping, we prepare an initial density profile with a Mott insulating core of one particle per site, see \figref{fig:DM-Timeevol-Mott}{c}. It is described by the local density matrix, $\rho_{\bf r}^{m,n}(0)=\delta_{m,n}\delta_{n,1}$. Since there is no condensate present, the symmetry-breaking terms like $\Phi_{\bf r}(t)$ and $G_{\bf r}^{(c)\alpha, \beta}(t)$ with $\alpha \neq \beta$ are zero, initially as well as during the time evolution, leading to a vanishing mean-field term $\hat{\widetilde{H}}\vphantom{\hat{H}}_{\rm S}^{\rm MF}(t)$ in \cref{eq:EOMrho0Markovian}. As a result, the time-evolved density matrices $\hat{\rho}_{\bf r}(t)$ at each site ${\bf r}$ remain diagonal and, therefore, commute with the local Hamiltonian $\hat{H}_{\rm S}$ leading the commutator term in \cref{eq:EOMrho0Markovian} to vanish. The expansion dynamics of all the non-condensed atoms in the Mott insulating state are, thus, governed by the diagonal correlator components $G_{\bf r}^{(c) \alpha,\alpha}(t)$. Thus, the local density-matrix is effectively time evolved by the master equation
\begin{align}
    \partial_t \hat{\rho}_{\bf r}(t) &= J^2 \tau_{\rm corr} \sum_{{\bf r'} \in {\bf R_r}} \sum_{\alpha} G^{(c)\alpha,\alpha}_{\bf r'}(t) \left( 2 b_{{\bf r}, \alpha}^\pdagger \hat{\rho}_{\bf r}(t) b^{\dagger}_{{\bf r},\alpha} \nonumber \right. \\
    &- \left. b^{\dagger}_{{\bf r}, \alpha} b_{{\bf r}, \alpha}^\pdagger \hat{\rho}_{\bf r}(t) - \hat{\rho}_{\bf r}(t) b^{\dagger}_{{\bf r}, \alpha} b_{{\bf r}, \alpha}^\pdagger \right) .
\end{align}
Thus, $J^2 \tau_{\rm corr}/U$ turns out to be the only relevant scale governing the transport of non-condensed particles without condensate. The corresponding expansion dynamics of non-condensed particles are demonstrated in \cref{fig:DM-Timeevol-Mott}. The long-time behavior shown by the spatio-temporal profile in \figref{fig:DM-Timeevol-Mott}{b} reveals a slow, non-ballistic expansion of the non-condensed bosons, and a circularly symmetric spatial profile is observed, see \figref{fig:DM-Timeevol-Mott}{c}.

\begin{figure}[t]
    \includegraphics[width=\linewidth]{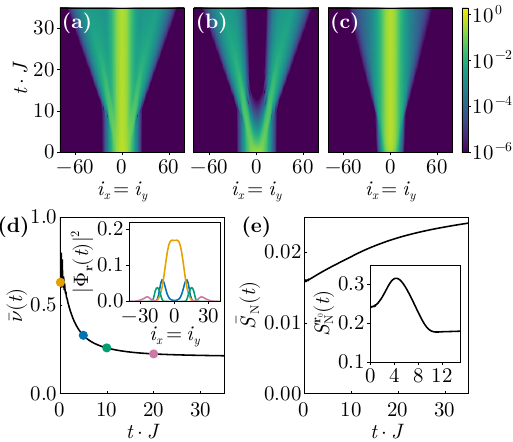}
    \caption{{\it Expansion dynamics of a condensate gas.} The spatio-temporal profiles of (a) the particle-number density $N_{\bf r}(t)$, (b) condensate density $|\Phi_{\bf r}(t)|^2$ and (c) non-condensed density $N_{\bf r}^{\rm fl}(t)$ along the diagonal cut of a square lattice are shown for $U/J=20$. The dynamics of the total condensate fraction $\bar{\nu}(t)$ are plotted in (d).  The inset shows the spatial profile of $|\Phi_{\bf r}(t)|^2$ along the diagonal at the time instants $t\cdot J=0.1$, $5$, $10$ and $20$, marked by the filled circles in the main panel using the same color code. (e) The time evolution of the number entropy averaged over the complete lattice, $\bar{S}_{\rm N}(t)$, and at the lattice center, ${\bf r}_\textrm{c} = (0,0)$, are shown in the main panel and in the inset, respectively.} 
    \label{fig:SF-expansion}
\end{figure}

To characterize the slow expansion, we compute the spatial width $\sigma_{\rm w}(t)$ of the expanded cloud from the Mott insulating state given by
\begin{equation}
    \sigma_{\rm w}(t)= \sqrt{\langle {\bf r}^2 \rangle - \langle {\bf r}\rangle ^2}, ~ \langle (\cdot)\rangle = \frac{1}{N_{\rm tot}}\sum_{\bf r} (\cdot) ~ N_{\bf r}(t) 
\end{equation}
In contrast to the ballistic expansion of the condensate, the non-condensed atoms in the Mott state expand diffusively, that is, $\sigma_{\rm w}(t) \propto \sqrt{t}$, see \figref{fig:S-and-width-Mott}{a}. The diffusive expansion reflects the incoherent nature of the noncondensed atoms in the Mott state, while the coherent condensate atoms \cite{Trujillo-Martinez_2021} as well as single particle-hole excitations (doublons or holons) out of a Mott state \cite{Ciuti20} can propagate ballistically. This contrasting behavior was also found in a classical vs. quantum random walk experiment \cite{Meschede15}. This diffusive transport of normal bosons can not be captured from the Gutzwiller mean-field method since it only allows hopping of the condensate, and not of the non-condensed particles. The diffusive expansion is corroborated by melting of the Mott state---the local Fock state probabilities are no longer delta functions as it was initially, but become distributions as shown for the center site ${\bf r}_{\rm c} = (0,0)$, see \figref{fig:S-and-width-Mott}{c}. Consequently, the number entropy $S_{\rm N}^{{\bf r}_{\textrm{c}}}(t)$ at ${\bf r}_{\rm c}$ first increases, but then decreases in the long time due to decrease in the local atom density $N_{{\bf r}_\textrm{c}}(t)$, see \figref{fig:S-and-width-Mott}{b} inset. The average number entropy $\bar{S}_{\rm N}(t)$, in contrast to the case with condensate, exhibits a very slow growth consistent with the diffusive expansion of non-condensed atoms as shown by \figref{fig:S-and-width-Mott}{b}.

\section{Dynamics of an initial superfluid condensate}
\label{sec:SF-expansion}

We now demonstrate the expansion of a superfluid condensate in the strongly correlated regime near the Mott insulator transition. To this end, we prepare an initial coherent state $|\Phi_{\bf r}\rangle$ with the local condensate amplitude $\Phi_{\bf r}=\sqrt{N_{\bf r}}$, that is, vanishing non-condensed density initially, and these condensate atoms are spatially distributed as a Gaussian given by
\begin{equation}
    N_{\bf r} = |\Phi_{\bf r}|^2 = \frac{V_0N_{\rm tot}}{J \pi} e^{-\frac{V_0 r^2}{J}}, ~ |\Phi_{\bf r}\rangle = e^{-\frac{|\Phi_{\bf r}|^2}{2}} \sum_{n=0}^{\infty} \frac{\alpha^n}{\sqrt{n!}} |n\rangle,
\end{equation}
with the spatial width $1/\sqrt{V_0/J}$ analogous to the ground state of a non-interacting Bose gas in a harmonic trapping potential, see \cref{fig:schematics} caption. The condensate cloud exhibits an expansion dynamics in the homogeneous lattice ($V_{\bf r}=0$) as shown in \cref{fig:SF-expansion}. In the transient times, the dynamics at the core are dominated by the interaction, and, thus, non-condensed atoms are generated leading to depletion of the condensate, see \figref{fig:SF-expansion}{b}. As a result, the condensate fraction $\bar{\nu}$ decreases at first as shown in \figref{fig:SF-expansion}{d}. The remaining condensate cloud expands ballistically, as can be observed from the wing of the expanding cloud in \figref{fig:SF-expansion}{b} and from the corresponding density profiles at different time instants shown in \figref{fig:SF-expansion}{d} inset. The non-condensed cloud generated from the condensate due to the interaction is either dragged along with the condensate, or expands diffusively (cf. \figref{fig:SF-expansion}{c}), see also \figref{fig:DM-Timeevol}{c} and the discussion in \cref{sec:Expansion-cond-NC}. As anticipated, the corresponding spatial profiles of all types of atom densities look qualitatively the same as in \figref{fig:GW-DM-Profiles}{a--c}, and, therefore, are not shown here. 

\section{Conclusions}
\label{sec:conclusion}

We have developed a theoretical method for the spatio-temporal evolution of ultracold Bose gases on large, two-dimensional lattices in the strongly interacting regime. Our method goes beyond the frequently used, time-dependent Gutzwiller theory \cite{Rigol_2011, Rigol17} and the cluster mean-field theory \cite{Ray24} in that it includes motion of the condensate amplitude and hopping of non-condensed fluctuations on the same footing and, thus allows to describe, for instance, the expansion from the Mott localized phase. Being based on a master equation for the density matrix, this method gives access to the dynamics of all single-time expectation values local in space, which are the most relevant quantities in gas-expansion situations. It allows to describe thermodynamically large systems and is numerically far less costly than a space-time dependent, bosonic dynamical mean-field theory \cite{Eckstein15}, applied to spatially inhomogeneous situations, would be. 

As an example, we have demonstrated the expansion dynamics of strongly interacting bosons in a two-dimensional Hubbard lattice. Starting from a realistic initial state of a Mott localized core surrounded by a condensed ring, we observed a fast, ballistically expanding halo of condensate and a slow, diffusively spreading core of non-condensed atoms, consistent with experiments Ref.~\cite{Ronzheimer_2013}. 
Thus, our approach can be applied in a straight-forward way to other strongly correlated Bose lattice problems, like the Dicke-Bose-Hubbard model \cite{Esslinger_2016, Ray24}, or the localization dynamics in higher-dimensional  disordered systems, including the Bose glass phase \cite{Schneider_2024}. 

\section*{Acknowledgments}

We sincerely thank Michael Kajan for useful discussions, especially during the early stage of this work. This work was funded by the Deutsche Forschungsgemeinschaft (DFG) under Germany’s Excellence Strategy-Cluster of Excellence Matter and Light for Quantum Computing, ML4Q (No. 390534769) and through the DFG Collaborative Research Center CRC 185 OSCAR (No. 277625399).\\

Data availability –- The data that support the findings of this article are openly available \cite{Zenodo-data_2025}.

\appendix

\section{Derivation of the quantum master equation}
\label{deriv-QME}

In this Appendix, we lay out the details of the derivations of the Master equation \cref{eq:EOMrho0Markovian}. We begin with the density matrix equation \cref{eq:EOMrho0-BHM} in interaction picture,
\begin{equation}
    \partial_t \hat{\rho}^I_{\bf r}(t) = - \int_{t_0}^t \mathrm{d}t' ~ \mathrm{Tr}_{({\bf r})}\left[\hat{H}_{\rm SB}^{{\rm fl},\,I}(t),\left[\hat{H}_{\rm SB}^{{\rm fl},\,I}(t'),\hat{\rho}^I(t')\right]\right].
    \label{Aeq:EOMrho0-BHM}
\end{equation}
The integrand in \cref{Aeq:EOMrho0-BHM} can be rearranged as follows:
\begin{align}
    &\left[\hat{H}_{\rm SB}^{{\rm fl},\,I}(t),\left[\hat{H}_{\rm SB}^{{\rm fl},\,I}(t'),\hat{\rho}^I(t')\right]\right] \nonumber \\
    = &\left[\hat{H}_{\rm SB}^{{\rm fl},\,I}(t),\hat{H}_{\rm SB}^{{\rm fl},\,I}(t') \hat{\rho}^I(t')\right] - \left[\hat{H}_{\rm SB}^{{\rm fl},\,I}(t),\hat{\rho}^I(t') \hat{H}_{\rm SB}^{{\rm fl},\,I}(t')\right] \nonumber \\
    = &\left[\hat{H}_{\rm SB}^{{\rm fl},\,I}(t),\hat{H}_{\rm SB}^{{\rm fl},\,I}(t') \hat{\rho}^I(t')\right] + \mathrm{H.c.}
    \label{Aeq:integrand}
\end{align}
From \cref{eq:H-hybrid} the third term $\hat{H}_{\rm SB}^{\rm fl}$ ($2^{\rm nd}$ order in fluctuations) can be written in the Nambu representation as
\begin{align}
     \hat{H}_{\rm SB}^{\rm fl} &= -J \sum_{{\bf r'} \in {\bf R_r}} \delta\hat{b}_{\bf r'}^\dagger \delta\hat{b}_{\bf r^\pprime}^\pdagger + {\rm h.c.} \nonumber \\
     &= -J \sum_{{\bf r'} \in {\bf R_r}} \sum_{\alpha} \delta b_{{\bf r'},\, \alpha}^\dagger \delta b_{{\bf r^\pprime},\, \alpha}^\pdagger \, .
     \label{Aeq:Hfl_SB}
\end{align}
Now, putting \cref{Aeq:Hfl_SB} into \cref{Aeq:integrand}, we obtain an explicit form of \cref{Aeq:EOMrho0-BHM} given by
\begin{align}
    \partial_t \hat{\rho}^I_{\bf r}(t) &= - J^2 \int_{t_0}^t \mathrm{d}t' \sum_{{\bf r'},{\bf r''} \in {\bf R_r}} \sum_{\alpha,\, \beta} \mathrm{Tr}_{({\bf r})} \left\{\left[\delta b_{{\bf r'},\, \alpha}^{\dagger \, I}(t) \right. \right. \nonumber \\
    &\times \left. \left. \delta b_{{\bf r^\pprime},\, \alpha}^{I\pdagger}(t), \left[\delta b_{{\bf r''},\, \beta}^{\dagger \, I}(t') \delta b_{{\bf r^\pprime},\, \beta}^{I\pdagger}(t'),\hat{\rho}^I(t')\right]\right]\right\}
\end{align}
Due to cyclic permutation of operators on sites ${\bf r'},{\bf r''} \neq {\bf r}$ under the trace, the density matrix equation in \cref{eq:EOMrho0-BHM} can be written in the Nambu representation as
\begin{align}
    \partial_t \hat{\rho}^I_{\bf r}(t) &= -J^2 \int_{t_0}^t \mathrm{d}\bar{t} \sum_{{\bf r'},{\bf r''} \in {\bf R_r}} \sum_{\alpha,\, \beta} \Big( \left[b^{\dagger \, I}_{{\bf r},\,\alpha}(t), b^{I}_{{\bf r},\, \beta}(\bar{t}) \hat{\rho}^I_{\bf r}(\bar{t})\right] \nonumber \\
    &- \varphi_{{\bf r^\pprime},\,\beta}(\bar{t}) \left[b^{\dagger \, I}_{{\bf r},\, \alpha}(t), \hat{\rho}^I_{\bf r}(\bar{t})\right] \Big) G^{(c) \, \alpha,\,\beta}_{{\bf r',r''}}(t,\bar{t}) + {\rm H.c.} \, ,
    \label{Aeq:ME-int-1}
\end{align}
which is \cref{eq:EOM-rho0-BHM_1} in the main text, where $G^{(c) \, \alpha,\,\beta}_{{\bf r',r''}}(t,\bar{t})$ is the two-time correlator, see \cref{eq:2-time-corr}. To arrive at \cref{Aeq:ME-int-1}, we have re-written the fluctuations at local site ${\bf r}$ as $\delta b_{{\bf r},\,\alpha}(t) = b_{{\bf r},\,\alpha} - \varphi_{{\bf r},\,\alpha}(t)$.
Using the approximation in \cref{eq:EqualTimeApprox} and the following paragraph in the main text, one is left with the following integral in the relative time $\tau=(t-\bar{t})$ of the Wigner coordinates: 
\begin{equation}
    \boldsymbol{G}_{{\bf r'},\,{\bf r''}}^{(c)}(T,\tau=0)\int_0^{\infty} d\tau~ e^{-\tau/\tau_{\rm corr}} \rightarrow \tau_{\rm corr} \boldsymbol{G}_{\bf r'}^{(c)}(t)\delta_{{\bf r'},\,{\bf r''}},
    \label{Aeq:time-loc}
\end{equation}
Note that, due to factorization of the density matrix into local sites, spatially non-local correlations must vanish at equal time. Thus, the correlator in \cref{eq:2-time-corr} reduces to the form, local in time and space,
\begin{equation}
    G_{\bf r'}^{(c) \, \alpha,\,\beta}(t) = \mathrm{Tr}\left\{ {\delta} b^{I}_{{\bf r'},\, \alpha}(t) \delta b^{\dagger\, I}_{{\bf r'},\, \beta}(t) \hat{\rho}_{{\bf r'}}^{I}(t) \right\}.  
    \label{Aeq:corr-equal-time}
\end{equation} 
Also, in \cref{Aeq:time-loc} the upper limit of $\tau$ is pushed to $\infty$ due to the fast decay of the exponential, see the discussion below \cref{eq:EqualTimeApprox}. The resulting equation of motion is, thus, local in time $t\equiv T$, and is given by
\begin{align}
    \partial_t \hat{\rho}^I_{\bf r}(t) &= -J^2 \tau_{\rm corr} \sum_{{\bf r'} \in {\bf R_r}} \sum_{\alpha,\, \beta} \left( \left[b^{\dagger \, I}_{{\bf r^\pprime},\,\alpha}(t), b^{I\pdagger}_{{\bf r},\, \beta}(t) \hat{\rho}^I_{\bf r}(t)\right] \right. \nonumber \\
    &- \left. \varphi_{{\bf r^\pprime},\,\beta}(t) \left[b^{\dagger \, I}_{{\bf r^\pprime},\, \alpha}(t), \hat{\rho}^I_{\bf r}(t)\right] \right) G^{(c) \, \alpha,\,\beta}_{{\bf r'}}(t) + {\rm H.c.}
    \label{Aeq:ME-int-2}
\end{align}
In this time-local form the transformation back to Schr{\"o}dinger picture is straightforward, and can be performed using $\hat{U}_S(t,t_0)$ (cf. \cref{eq:Unitary-interaction}), which yields, for the left hand side of \cref{Aeq:ME-int-2},
\begin{align}
    &\hat{U}_S(t,t_0) \left[\partial_t \hat{\rho}^{I}_{\bf r}(t)\right] \hat{U}^\dagger_S(t,t_0) \nonumber \\
    =& \partial_t \hat{\rho}_{\bf r}(t) 
    + i\left[ \hat{H}_S + \hat{H}_S^{\mathrm{MF}}(t), \hat{\rho}_{\bf r}(t) \right] .
\end{align}
Next, we evaluate the right hand side of \cref{Aeq:ME-int-2}. First, we note that the correlator in \cref{Aeq:corr-equal-time}, in the Schr{\"o}dinger picture, yields
\begin{align}
    G_{\bf r'}^{(c) \alpha,\, \beta}(t) =& \mathrm{Tr}\left\{ {\delta} b_{{\bf r'},\, \alpha}^\pdagger(t) \delta b^{\dagger}_{{\bf r'},\, \beta}(t) \hat{\rho}_{{\bf r}}(t) \right\}  \\
    =& \mathrm{Tr}\left\{ b_{{\bf r'},\, \alpha}^\pdagger b^{\dagger}_{{\bf r'},\, \beta} \hat{\rho}_{{\bf r'}}(t) \right\} - \varphi_{{\bf r'},\, \alpha}^{\vphantom{*}}(t) \varphi^*_{{\bf r'},\, \beta}(t) \, , \nonumber
\end{align}
which is \cref{eq:Eq-time-corr} in the main text. 
Using the symmetry properties of $G^{(c) \, \alpha,\,\beta}_{{\bf r'}}(t)$ with respect to its Nambu indices, namely
\begin{equation}
    \left(G^{(c) \alpha,\,\beta}_{{\bf r'}}(t)\right)^* = G^{(c) \beta, \, \alpha}_{{\bf r'}}(t) \, ,
\end{equation}
we, first, explicitly compute the terms which do not contain the condensate amplitude $\varphi_{{\bf r},\,\beta}(t)$:
\begin{widetext}
\begin{align}
    &\hat{U}_S(t,t_0) \sum_{{\bf r'} \in {\bf R_r}} \sum_{\alpha,\, \beta} \left[b^{\dagger \, I}_{{\bf r^\pprime},\,\alpha}(t), b^{I\pdagger}_{{\bf r},\, \beta}(t) \hat{\rho}^I_{\bf r}(t)\right] \times G^{(c)\alpha,\beta}_{\bf r'}(t)  ~\hat{U}^\dagger_S(t,t_0) + \text{H.c.} \nonumber \\
    =& \sum_{{\bf r'} \in {\bf R_r}} \sum_{\alpha,\, \beta} G^{(c)\alpha,\beta}_{\bf r'}(t) \left[b^{\dagger}_{{\bf r^\pprime},\,\alpha}(t), b_{{\bf r},\, \beta}^\pdagger(t) \hat{\rho}_{\bf r}(t)\right] + \text{H.c.} \nonumber \\
    =& \sum_{{\bf r'} \in {\bf R_r}} \sum_{\alpha,\beta} G^{(c)\alpha,\beta}_{\bf r'}(t) \left( 2 b_{{\bf r}, \beta}^\pdagger \hat{\rho}_{\bf r}(t) b^{\dagger}_{{\bf r^\pprime},\alpha} - b^{\dagger}_{{\bf r^\pprime}, \alpha} b_{{\bf r}, \beta}^\pdagger \hat{\rho}_{\bf r}(t) - \hat{\rho}_{\bf r}(t) b^{\dagger}_{{\bf r^\pprime}, \alpha} b_{{\bf r}, \beta}^\pdagger \right),
\end{align}
where the expression in the last line is $\mathcal{L}[\boldsymbol{b}_{\bf r}^\pdagger]$ in \cref{eq:D-fluctuation}. Finally, the terms containing the condensate amplitudes $\varphi_{{\bf r},\,\beta}(t)$ can be written in the Schr{\"o}dinger picture as
\begin{align}
    &\hat{U}_S(t,t_0) \sum_{{\bf r'} \in {\bf R_r}} \sum_{\alpha,\, \beta} \varphi_{{\bf r}}^{\beta}(t) \left[b^{\dagger \, I}_{{\bf r},\, \alpha}(t), \hat{\rho}^I_{\bf r}(t)\right] G^{(c)\alpha,\beta}_{\bf r'}(t)  ~\hat{U}^\dagger_S(t,t_0) + \text{H.c.} \nonumber \\
    =& \sum_{\alpha,\, \beta} G^{(c)\alpha,\beta}_{\bf r'}(t) \left( \left[\varphi_{{\bf r},\,\beta}(t) b^{\dagger}_{{\bf r},\, \alpha}, \hat{\rho}_{\bf r}(t)\right] - \left[\varphi_{{\bf r},\,\alpha}^{*}(t) b_{{\bf r},\, \beta}, \hat{\rho}_{\bf r}(t)\right]\right) \nonumber \\
    =& \left[ \Phi_{{\bf r}}^\pdagger(t) \hat{b}^{\dagger}_{{\bf r}} - \Phi_{{\bf r}}^{*\pdagger}(t) \hat{b}_{{\bf r}}^\pdagger, \hat{\rho}_{\bf r}(t)\right],
\end{align}
\end{widetext}
where the terms with $\alpha \neq \beta$ drop out. To arrive at the last line, we have used the bosonic commutation relation. This is the Lamb shift term that modifies the mean-field Hamiltonian $\hat{H}_{\rm S}^{\rm MF}(t)$ in \cref{eq:Lamb-Shift}. Collecting all the components the Master equation for $\hat{\rho}_{\bf r}(t)$ is given by
\begin{equation}
    \partial_t \hat{\rho}_{\bf r}(t) = 
    - i [\hat{H}_{\rm S} + \hat{\widetilde{H}}\vphantom{\hat{H}}_{\rm S}^{\rm MF}(t), \hat{\rho}_{\bf r}(t)] + J^2 \tau_{\rm corr} \mathcal{L}[\boldsymbol{b}_{\bf r}^\pdagger],
    \label{eq:EOMrho0Markovian-Appen}
\end{equation}
which is Eq.~\eqref{eq:EOMrho0Markovian} in the main text. The form of the modified mean-field Hamiltonian $\hat{\widetilde{H}}\vphantom{\hat{H}}_{\rm S}^{\rm MF}(t)$ and the fluctuation term $\mathcal{L}[\boldsymbol{b}_{\bf r}^\pdagger]$ are given in \cref{eq:Lamb-Shift} and \cref{eq:D-fluctuation}, respectively. This completes our derivation of the master equation in \cref{eq:EOMrho0Markovian} used for the time evolution of local density matrices at each lattice site. Note that for a $L\times L$ square lattice, Eq.~\eqref{eq:EOMrho0Markovian-Appen} constitutes a closed set of $L^2$ coupled, first-order-in-time differential equations. Their numerical time evolution provides the time evolved density matrix $\hat{\rho}_{\bf r}(t)$, which is used to compute the observables local in space and time, such as particle number, condensate density and non-condensed fluctuations, as well as number entropy. These local quantities are relevant for cold atom experiments, and are presented and discussed as main results in the main text.   

\begin{figure}[t]
    \includegraphics[width=\linewidth]{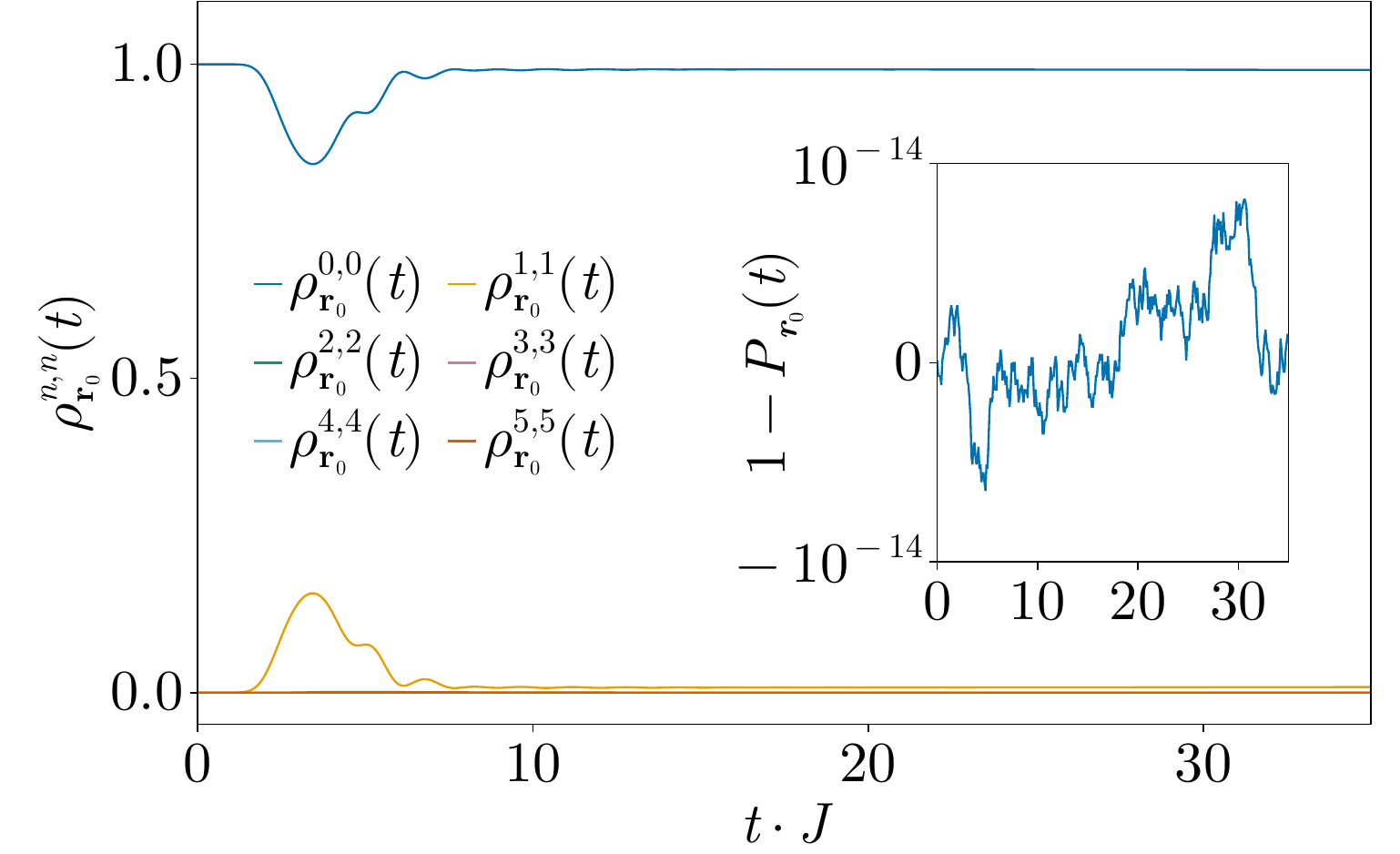}
    \caption{The time evolution of the Fock state probabilities $\rho_{{\bf r}_0}^{n,n}(t)$ up to $n=5$ at the site ${\bf r}_0$ corresponding to \cref{fig:Evolution-local} are shown. The inset shows the conservation of total Fock state probability $P_{{\bf r}_0}(t)=1$.}
    \label{fig:Fock-prob-dynamics}
\end{figure}

\section{Dynamics of local Fock state probability}
\label{app:B}

In Fig.~\ref{fig:Fock-prob-dynamics}, we show the time-evolution of the Fock state probabilities $\rho_{{\bf r}_0}^{n,n}(t)$ at the site ${\bf r}_0$ corresponding to \cref{fig:Evolution-local}. The chosen site ${\bf r}_0$ is initially in a vacuum state by preparation, that is, $\rho_{{\bf r}_0}^{n,m}(0)=\delta_{n,m}\delta_{n,0}$, and only gets excited as the superfluid hits the site during the ballistic expansion, leading to finite probabilities of higher Fock states, especially, $\rho_{{\bf r}_0}^{1,1}(t)$. The Fock state distributions at different time instants are shown in \figref{fig:Evolution-local}{c}. We note that by property of the master equation in \cref{eq:EOMrho0Markovian}, the total probability in every single site is conserved, see, for instance, the  total probability $P_{{\bf r}_0}(t)$ at the site ${\bf r}_0$ in the inset of \cref{fig:Fock-prob-dynamics}.

\bibliography{references}

\end{document}